\documentclass[prb,twocolumn,showpacs,superscriptaddress]{revtex4}

\usepackage{graphicx}
\usepackage{dcolumn}
\usepackage{amsmath}
\usepackage{color}

\newcommand{\etal}{\textit{et al.~}}

\begin{document}
\title{Structural and magnetic dimers in the spin-gapped system CuTe$_2$O$_5$}

\author{J.~Deisenhofer}
\affiliation{D\'{e}partement de Physique de la Mati\`{e}re Condens\'{e}e,
Universit\'{e} de Gen\`{e}ve, CH-1211 Gen\`{e}ve 4, Switzerland}

\author{R.~M.~Eremina}
\affiliation{EP V, Center for Electronic Correlations and Magnetism,
Augsburg University, D-86135 Augsburg, Germany}
\affiliation{E.~K.~Zavoisky Physical-Technical Institute, 420029
Kazan, Russia}

\author{A.~Pimenov}
\affiliation{EP V, Center for Electronic Correlations and Magnetism,
Augsburg University, D-86135 Augsburg, Germany}

\author{T.~Gavrilova}
\affiliation{E.~K.~Zavoisky Physical-Technical Institute, 420029
Kazan, Russia}

\author{H.~Berger}
\affiliation{Institute de Physique de la Mati\`{e}re Complexe, EPFL,
CH-1015 Lausanne, Switzerland}

\author{M. Johnsson}
\affiliation{Department of Inorganic Chemistry, Stockholm Univ.,
S-10691 Stockholm, Sweden}

\author{P.~Lemmens}
\affiliation{Institute for Physics of Condensed Matter, Technical
University of Braunschweig, D-38106 Braunschweig, Germany}

\author{H.-A.~Krug von Nidda}
\affiliation{EP V, Center for Electronic Correlations and Magnetism,
 Augsburg University, D-86135 Augsburg,
Germany}

\author{A.~Loidl}
\affiliation{EP V, Center for Electronic Correlations and Magnetism,
 Augsburg University, D-86135 Augsburg,
Germany}

\author{K.-S. Lee}
\affiliation{Department of Chemistry, North Carolina State
University, Raleigh, NC 27695-8204, USA}

\affiliation{Department of Chemistry, The Catholic University of
Korea, Bucheon, Gyeonggi-Do, South Korea 422-743}

\author{M.-H.~Whangbo}
\affiliation{Department of Chemistry, North Carolina State
University, Raleigh, NC 27695-8204, USA}

\date{\today}

\begin{abstract}
We investigated the magnetic properties of the system CuTe$_2$O$_5$
by susceptibility and electron spin resonance measurements. The
anisotropy of the effective $g$-factors and the ESR linewidth
indicates that the anticipated structural dimer does not correspond
to the singlet-forming magnetic dimer. Moreover, the spin
susceptibility of CuTe$_2$O$_5$ can only be described by taking into
account interdimer interactions of the same order of magnitude than
the intradimer coupling. Analyzing the exchange couplings in the
system we identify the strongest magnetic coupling between two Cu
ions to be mediated by super-super exchange interaction via a
bridging Te ligand, while the superexchange coupling between the Cu
ions of the structural dimer only results in the second strongest
coupling.
\end{abstract}



\maketitle

\section{Introduction}

Transition-metal-compounds based on Cu$^{2+}$ ions with a 3$d^9$
configuration exhibit an enormously rich variety of magnetic
structures depending on the effective magnetic dimensionality of the
system.\cite{Lemmens03} Introducing lone-pair cations like Se$^{4+}$
or Te$^{4+}$ into the magnetic system was suggested as a fruitful
path to tailor the magnetic dimensionality and to create new
magnetic structures.\cite{Johnsson00,Herak05} For example, it was
proposed that Cu$_2$Te$_2$O$_5$X$_2$ (X=Cl, Br), which consists of
tetrahedral clusters of Cu$^{2+}$ linked by bridging Te-O units, is
an example of quasi-zero-dimensional systems in which the extreme
limits of magnetic insulation via Te$^{4+}$ ions are
reached.\cite{Lemmens01} The transition from a spin-gapped
paramagnetic state to an antiferromagnetically ordered state at
$T_N=18.2$~K (X=Cl) and $T_N=11.4$~K (X=Br) was attributed to the
proximity of a quantum phase transition.\cite{Gros03} The
magnetization, specific heat and Raman scattering data of
Cu$_2$Te$_2$O$_5$Br$_2$ were also interpreted by considering weakly
coupled Cu$_4$ tetrahedra within a mean field
approximation.\cite{Gros03} Recently, this understanding in terms of
weakly-interacting tetramers has been questioned by a spin dimer
analysis, which showed that the spin exchange interactions between
adjacent Cu$^{2+}$ sites are much weaker within each Cu$_4$
tetrahedron than between adjacent Cu$_4$
tetrahedra.\cite{Whangbo03a} Additionally, electronic band structure
calculations for Cu$_2$Te$_2$O$_5$X$_2$ (X=Cl, Br) have confirmed
the presence of strong interactions between adjacent Cu$_4$
tetrahedra.\cite{valenti03}

\begin{figure}[b]
\centering
\includegraphics[width=60mm,clip,angle=0]{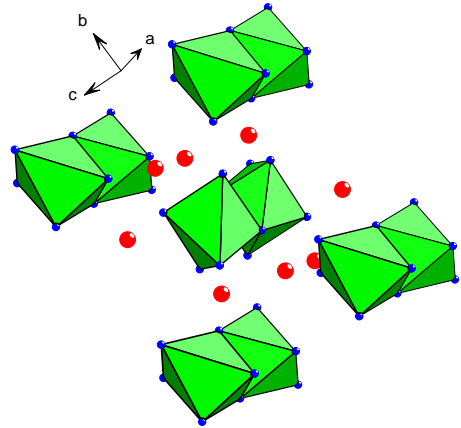}
\vspace{2mm} \caption[]{\label{struktur} (Color online) Crystal
structure of CuTe$_2$O$_5$~(space group $P21$/c). The Cu$_2$O$_{10}$
dimer units (edge-sharing octahedra) are separated by Te ions (large
spheres).}
\end{figure}

The compound investigated in this study is the related system
CuTe$_2$O$_5$ which exhibits a monoclinic structure with space group
P21/c and lattice parameters $a=6.871$\AA, $b=9.322$\AA,
$c=7.602$\AA, $\beta=109.08^\circ$.\cite{Hanke73} The lattice
consists of pairs of strongly distorted and edge-sharing CuO$_6$
octahedra with a Cu-Cu distance of 3.18 {\AA}. These structural dimer
units (see Fig.~\ref{struktur}) are separated by Te-O bridging
ligands and a Cu-Cu distance of 5.28~{\AA}. The magnetic susceptibility
of CuTe$_2$O$_5$~shows a maximum at $T_{\rm max}=56.6$~K and a
strong decrease for lower temperatures, which can be roughly modeled
by isolated magnetic dimers. The high-temperature susceptibility
corresponds to a Curie-Weiss law with a Curie-Weiss temperature of
$\Theta=-41$~K.\cite{Lemmens03} In this study we investigate the
spin susceptibility of CuTe$_2$O$_5$ by electron spin resonance
(ESR) and dc-susceptibility measurements. By analyzing the spin
susceptibility and the exchange paths within and between the
structural dimers based on extended H\"{u}ckel tight binding (EHTB)
electronic structure calculations, we determine the magnetic
structure of the system and find that it differs significantly from
the one intuitively imposed by the lattice structure.

\section{Sample preparation and experimental details}

Large single crystals of CuTe$_2$O$_5$ in form of platelets with a
maximum size of 8x8x0.5 mm$^3$ were grown by the usual halogen vapor
transport technique, using HBr as transport agents. The charge and
growth-zone temperatures were 580~$^\circ$C and 450~$^\circ$C
respectively. The stoichiometry of obtained single crystals was
quantitatively probed by electron-probe microanalysis. The compound
crystallizes as large blue-green plates and was characterized also
by X-ray diffraction. Powder samples have been prepared in sealed
quartz ampules using stoichiometric molar ratios of CuO and
TeO$_2$.\\ Susceptibility measurements were performed on single
crystals and polycrystalline samples using a SQUID magnetometer
(Quantum Design). ESR measurements were performed in a Bruker
ELEXSYS E500 CW-spectrometer at X-band (9.47 GHz) and Q-band (34
GHz) frequencies equipped with continuous He-gas-flow cryostats
(Oxford Instruments) in the temperature range $4.2 \leq T \leq
300$~K. ESR detects the power $P$ absorbed by the sample from the
transverse magnetic microwave field as a function of the static
magnetic field $H$. The signal-to-noise ratio of the spectra is
improved by recording the derivative $dP/dH$ using lock-in technique
with field modulation. The CuTe$_2$O$_5$~ single crystal was glued
on a suprasil-quartz rod, which allowed the rotation of the sample
around defined crystallographic axes. High-field ESR was performed
at a frequency of 185~GHz using a quasi optical technique with
backward-wave oscillators as coherent sources for submillimeter
wavelength radiation.\cite{Kozlov1998,Ivannikov02} The spectrometer
is equipped with a superconducting split-coil magnet with a maximum
field of 8T.

\section{Experimental Results}

\begin{figure}[h]
\centering
\includegraphics[width=70mm,clip,angle=0]{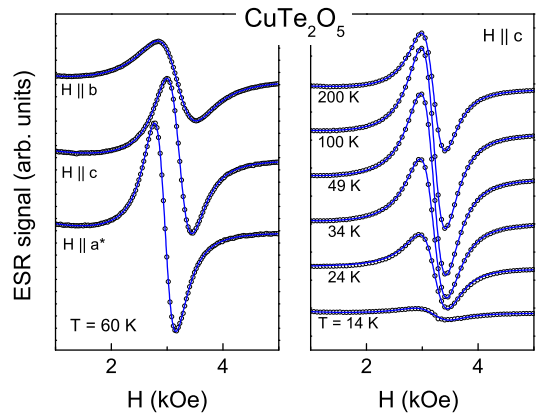}
\vspace{2mm} \caption[]{\label{spectra} (Color online) ESR spectra
in CuTe$_2$O$_5$~ at $T=60$ K for the external magnetic field
parallel to the main crystallographic directions (left column) and
at different temperatures for $\mathbf{H}\parallel c$ (right
column). The solid lines represent fits with a Lorentzian line
shape.}
\end{figure}

As shown in Fig.~\ref{spectra} the observed ESR absorption is well
described by a single exchange narrowed Lorentzian line with
resonance field $H_{\rm res}$ and half-width at half maximum
linewidth $\Delta H$ except for temperatures below 20~K, where the
ESR intensity $I_{\rm ESR}$ strongly decreases due to the singlet
formation of the dimers and the residual signal remains from
defects giving rise to a complicated splitting of the spectrum
like e.g. in CuGeO$_3$ below the spin-Peierls
transition.\cite{Smirnov1998}

Therefore, we will concentrate in the following on the temperature
regime above 20~K. In the investigated temperature regime the $g$
values are practically temperature independent with $g_{a^*} =
2.27(2)$, $g_{b} = 2.11(3)$, and $g_{c} = 2.10(2)$, where $a^*$
denotes the direction perpendicular to the $b$-$c$~plane. Such $g$
values slightly enhanced with respect to the spin-only value of
$g=2$ are typical for the $3d^9$ electronic configuration of
Cu$^{2+}$ in distorted oxygen octahedra, where the orbital momentum
is nearly quenched by the crystal field.\cite{Abragam1970}

\begin{figure}[t]
\centering
\includegraphics[width=80mm,clip,angle=0]{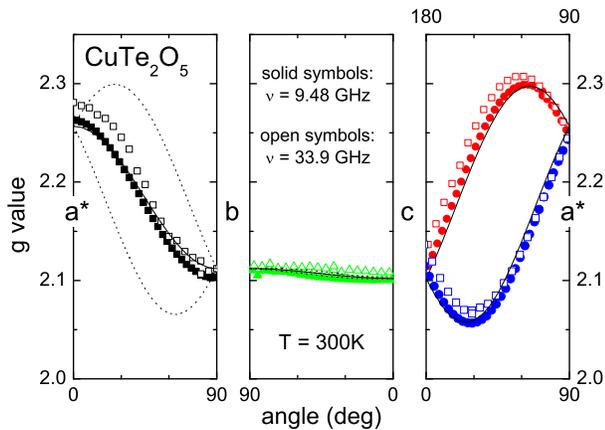}
\vspace{2mm} \caption[]{\label{grot300Kfit} (Color online) Angular
dependence of the $g$ factor for three perpendicular
crystallographic planes at room temperature for X-band and Q-band
frequencies. The solid lines represent a fit of the g tensor
assuming two inequivalent Cu sites, the contributions of which are
indicated by dotted lines. (Right frame: bottom abscissa: upper data
set; top abscissa: lower data set.)}
\end{figure}

Figure~\ref{grot300Kfit} shows the full angular dependence of the
$g$ tensor in three perpendicular crystallographic planes at room
temperature. It is nearly constant in the $b$-$c$~plane and
increases monotonously from $b$ to $a^*$. But within the
$a^*$-$c$~plane the principal axes of the tensor turn out to be
tilted by about $30^{\circ}$ with respect to the $a^*$ axis. Here
the $g$ tensor exhibits its maximum at $g_{\rm max}=2.30(2)$ and
minimum at $g_{\rm min}=2.06(2)$. The $a^*$-$c$~plane contains the
real crystallographic $a$ axis at the monoclinic angle
$\beta=109.08^{\circ}$, but this does not coincide with one of the
principal axes of the g tensor.

\begin{figure}[h]
\centering
\includegraphics[width=80mm,clip,angle=0]{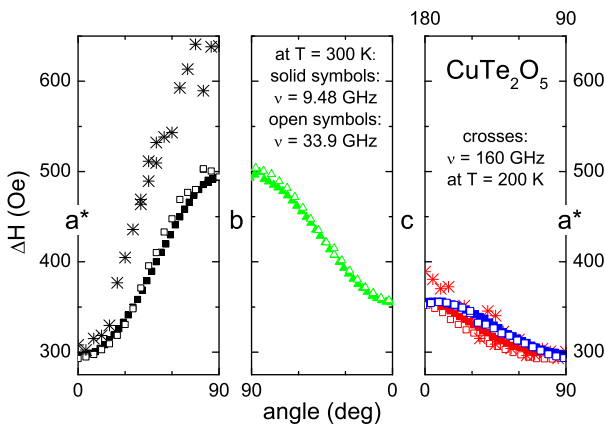}
\vspace{2mm} \caption[]{\label{dhrot300K} (Color online) Angular
dependence of the linewidth for three perpendicular crystallographic
planes at 300~K for X-band  and Q-band frequencies, and for the
$a^*-b$- and the $a^*-c$-plane (spheres) measured at 200~K and
160~GHz. (Right frame: bottom abscissa: lower data and crosses; top
abscissa: upper data)}
\end{figure}

\begin{figure}[b]
\centering
\includegraphics[width=80mm,clip,angle=0]{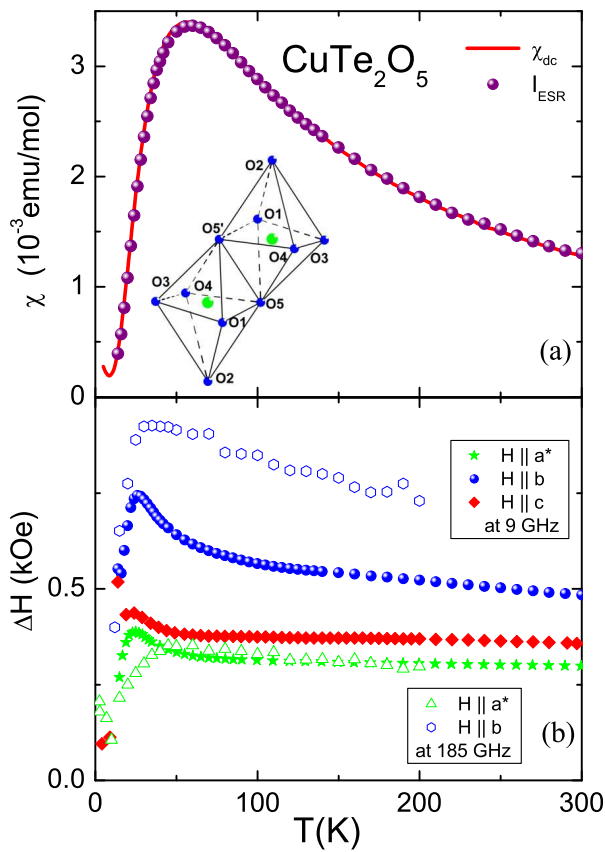}
\vspace{2mm} \caption[]{\label{ESRparameters} (Color online)
Temperature dependences of (a) the ESR intensity and the dc
susceptibility (Inset: Cu$_2$O$_10$ structural dimer), (b) the ESR
linewidth in CuTe$_2$O$_5$ with the external magnetic field applied
along the crystallographic axes for 9~GHz and 185~GHz.}
\end{figure}

The corresponding angular dependence of the linewidth at 300~K is
depicted in Fig.~\ref{dhrot300K}. The data approximately coincide
for X-band and Q-band frequency. The largest variation of the
linewidth appears in the $a^*$-$b$~plane where it monotonously
increases from 300~Oe up to 500~Oe on rotating the magnetic field
from the $a^*$ direction into the $b$ direction. On further rotating
the field into the $c$ direction the linewidth is monotonously
reduced to 350~Oe. Again, in the $a^*$-$c$~plane the extrema of the
linewidth do not coincide with $a^*$ and $c$ direction, but are
tilted by about $10^{\circ}$ with respect to these directions. The
temperature dependence of the linewidth is not very pronounced down
to about 100~K. As shown in Fig.~\ref{ESRparameters}(b), the
linewidth increases with decreasing temperature up to about 750~Oe
at 25~K for $H
\parallel b$ followed by a decrease to lower temperatures, whereas
the increase is much weaker for the two other orientations.

The temperature dependence of the ESR intensity obtained by double
integration of the ESR signal perfectly coincides with the
temperature behavior of the dc-susceptibility (see
Fig.~\ref{ESRparameters}(a)). Thus, the static susceptibility indeed
is dominated by the pure spin susceptibility represented by the ESR
intensity - diamagnetic and van Vleck contributions are negligible.
Before we analyze the spin susceptibility in detail in the next
section, we want to discuss the difficulties occurring in the
evaluation of $g$ tensor and linewidth, if the system is assumed to
consist of nearly isolated magnetic dimers as suggested from the
crystal structure.

We try to describe the angular dependence of the $g$ tensor using
purely structural considerations. Each Cu$^{2+}$ ion is surrounded
by an oxygen octahedron which is elongated along the
O2-O5$^{\prime}$ axis (see inset of Fig.~\ref{ESRparameters}(a)),
with distances Cu-O5$^{\prime}$=2.303~\AA~and Cu-O2=2.779~\AA. But
the Cu-O distances within the plane, built up by the other four
oxygen ions, differ only within 0.04~\AA, with the smallest
distances for Cu-O5 of 1.948~\AA, and Cu-O3 of
1.969~\AA.\cite{Hanke73} Note that because of the asymmetric
elongation of the octahedra, the ground-state orbital will have a
small admixture of the $d_{3z^2-r^2}$-orbital. As an approximation,
however, we assume that also in the present geometry the hole of the
$3d^9$ state occupies the $d_{x^2-y^2}$ orbital located in this
plane. This gives rise to a $g$ tensor diagonal in a local cartesian
coordinate system with two axes approximately given by Cu-O1 ($x$)
and Cu-O5 ($y$) and the $z$ axis (with the largest $g$ value)
perpendicular to the plane. The oxygen octahedra of two
Cu$^{2+}$-ions within one dimer are just rotated by $180^{\circ}$
against each other and, therefore, should have the same $g$-tensor.
Hence, to describe the anisotropy of the $g$-value, one has to sum
up the local $g$-tensors of the two dimers with different
orientation in the unit cell. The best fit yielding $g_{xx}=2.029$,
$g_{yy}=2.105$, and $g_{zz} = 2.329$ is shown in
Fig.~\ref{grot300Kfit} as solid lines together with the single
contributions (dotted lines) of the two dimers of different
orientation. The strongest effect of the two inequivalent places is
visible in the $a^{*}$-$b$~plane, whereas it is only weak in the
$b$-$c$~plane and vanishes completely in the $a^{*}$-$c$~plane. In
the dimer model this should influence the line broadening in the
$a^*$-$b$~plane via the anisotropic Zeeman effect as will be
illustrated in the following.

In general, the ESR linewidth in the case of sufficiently strong
exchange interaction can be analyzed in terms of the
high-temperature Kubo-Tomita approach:\cite{Kubo54}
\begin{equation} \label{Dhkubo}
\Delta H_\infty = \frac{1}{g \mu_{\textrm{B}}} \cdot \frac{M_2}{J},
\end{equation}
where $\Delta H_\infty$ corresponds to the high-temperature value
of the ESR linewidth, $\mu_{\textrm{B}}$ denotes the Bohr
magneton, $J$ is the isotropic exchange constant, and $M_2$ yields
the second moment
\begin{equation}\label{2Mom}
M_2 = h^2 \langle(\nu - \nu_0)^2\rangle =  \frac{\langle [ {\cal
H}_{\textrm{int}}, S^+] \, [S^-, {\cal H}_{\textrm{int}}] \rangle}
{\langle S^+ S^- \rangle}
\end{equation}
of the ESR absorption line. It has been shown for several
antiferromagnetic spin $S=1/2$ chain compounds (e.g. LiCuVO$_4$ and
CuGeO$_3$) with a similarly large linewidth like CuTe$_2$O$_5$ that
the anisotropic exchange interactions are the dominant contributions
to the microscopic spin Hamiltonian ${\cal
H}_{\textrm{int}}$.\cite{KvN2002,Eremina2003,Eremin06,Zakharov06}
Therefore, we believe that the ESR linewidth in CuTe$_2$O$_5$ is
dominated by the anisotropic exchange, as well. Besides this
broadening mechanism, the anisotropic Zeeman interaction mentioned
above is expected to be of importance in the $a^*$-$b$~plane, where
the difference of the $g$ values of both inequivalent Cu sites
yields a contribution to the second moment which is proportional to
the square of the applied magnetic field . Details are described by
Pilawa for the case of CuGeO$_3$, where a similar situation occurs
due to two inequivalent Cu chains.\cite{Pilawa1997} Assuming that
the inter-dimer isotropic exchange interaction, which is responsible
for the narrowing of the anisotropic Zeeman contribution, is much
weaker than the intra-dimer isotropic exchange interaction, this
contribution to the line broadening should be largest for an angle
of $45^{\circ}$ in the $a^*$-$b$~plane, where the difference in $g$
values of both sites is largest, but should vanish for $0^{\circ}$
and $90^{\circ}$, where the $g$ values are equal. Moreover, this
contribution should gain importance with increasing frequency due to
its quadratic field dependence.

However, such an anisotropic Zeeman contribution to the linewidth
cannot be detected at X-band and Q-band frequencies (c.f.
Fig.~\ref{dhrot300K}), where the data approximately coincide within
the experimental uncertainty. Therefore, we performed additional
high-field ESR measurements at 160~GHz and at 185~GHz. Having
investigated the orientation dependence of the linewidth at 200~K
and 160~GHz (see Fig.\ref{dhrot300K}), we found that a strong
increase of the linewidth is observed in the $a^*$-$b$~plane, but
changes within the $a^*$-$c$~plane are very small as compared to the
X/Q-band data and remain within the experimental uncertainty,
although one may want to infer a slight tendency to an additional
line broadening along the $c$-axis, which would be in agreement with
the expectation for an anisotropic Zeeman contribution.
Consequently, we measured the temperature dependence of the
linewidth with the magnetic field oriented along the $a^*$- and the
$b$~axis (shown together with the X-Band data in
Fig.~\ref{ESRparameters}(b)) at a higher frequency of 185~GHz and a
correspondingly higher magnetic field. Note that the increase is
maximal for the magnetic field applied along the $b$ direction in
contrast to the expectation resulting from the analysis of the $g$
tensor. This discrepancy indicates that the assumed dimer model,
which is based on the structural Cu$_2$O$_{10}$ units, does not
properly describe the magnetic properties of CuTe$_2$O$_5$. Maybe
also the $g$ tensors of the inequivalent Cu sites deviate from the
orientation suggested by the structural data and a possible
admixture of the $d_{3z^2-r^2}$-orbital will give important
contributions and has to be taken into account in order to obtain a
consistent description oft he ESR data. Therefore, a detailed
analysis of the microscopic exchange paths (as will be presented in
Section IV.B) has to be performed before one can try to understand
the ESR behavior in this compound.

\begin{figure}[t]
\centering
\includegraphics[width=80mm,clip,angle=0]{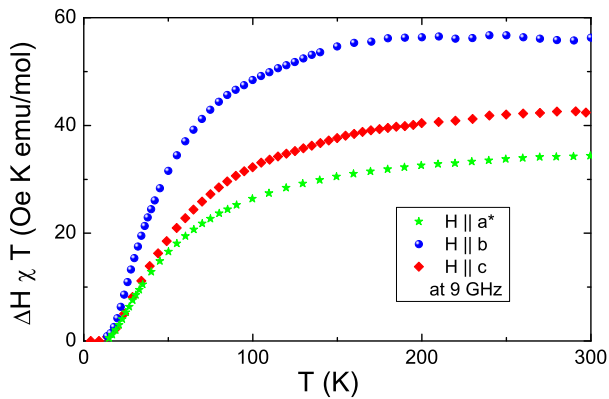}
\vspace{2mm} \caption[]{\label{memory} (Color online) Temperature
dependences of the memory function in CuTe$_2$O$_5$ with the
external magnetic field applied along the crystallographic axes.}
\end{figure}

Finally, we briefly comment on the temperature dependence of the
linewidth: Concerning the maximum of the linewidth at low
temperatures, we plot the product of the linewidth data with the
spin susceptibility multiplied by the temperature in
Fig.~\ref{memory}. In the Kubo-Tomita approach the main temperature
dependence of the ESR linewidth stems from the spin susceptibility:
\begin{equation} \label{Dhkubotemp}
\Delta H_{KT}(T) = \frac{1}{\chi(T)T} \cdot \Delta H_\infty,
\end{equation}
In this limit the product $\Delta H\cdot \chi\cdot T$ should
become temperature independent for high temperatures and
corresponds to the so called memory function. As recently reported
by Chabre {\it et al.} \cite{Chabre2005} for the quasi one
dimensional spin-gap system $\eta$-Na$_{1.286}$V$_2$O$_5$ the
maxima in the temperature dependence of the linewidth
(Fig.~\ref{ESRparameters}(b)) may just result from a monotonously
increasing memory function in combination with the temperature
dependence of the susceptibility. A similar temperature dependence
of the memory function has been experimentally found and
theoretically confirmed earlier by Pilawa {\it et al.} in the
one-dimensional antiferromagnet
tetraphenylverdazyl.\cite{Pilawa1995} The memory function in our
case behaves very similar to the above spin-chain compounds
starting from zero at zero temperature with a monotonous increase
and saturating at temperatures large compared to the exchange
coupling $J/k_{\rm B}$. Similar to the case of
$\eta$-Na$_{1.286}$V$_2$O$_5$ the anomaly suggested by the maximum
of the linewidth is not visible in the memory function any more
and, therefore, we discard the possibility that the maximum arises
due to changes in the magnetic structure of the system.

\section{Discussion}

\subsection{Fit of the spin susceptibility}

The usual ansatz to describe the spin susceptibility in dimer
systems is
\begin{equation}
 \chi=\chi_{VV} + \chi_C + \chi_{BB},
\end{equation}
where $\chi_{VV}$ denotes a possible Van-Vleck contribution,
$\chi_C=C/T$ is a Curie contribution due to unbound spins and
magnetic impurities, and $\chi_{BB}$ is the dimer susceptibility as
derived by Bleaney and Bowers (BB)\cite{Bleaney52}:
\begin{equation}
 \chi_{BB}(T)=\frac{Ng^2\mu_B^2}{k_B T}[3+ \exp{(J/k_BT)}]^{-1},
\end{equation}
where $J$ denotes the intradimer exchange coupling, $g$ is the
effective $g$-factor, and $\mu_B$ is the Bohr magneton.

To account for possible couplings between the dimers we will take
into account two different approaches. The first one is to
substitute the BB-equation by the approach by Johnston \etal who
calculated the susceptibility of a spin-1/2 antiferromagnetic chain
with alternating exchange constants $J_1$ and
$J_2$:\cite{Johnston00}
\begin{equation} \label{Johnston}
 \chi(\alpha,T)=\frac{Ng^2\mu_B^2}{k_B T}\chi^*(\alpha,T),
\end{equation}
where $\alpha=J_2/J_1$ denotes the ratio of the antiferromagnetic
coupling constants. The function $\chi^*(\alpha,T)$ has been
obtained by various numerical methods to describe excellently the
full range of susceptibilities from a pure Heisenberg dimer system
with $\alpha=0$ (corresponding to the Bleaney-Bowers equation) to a
uniform Heisenberg spin chain with $\alpha=1$. Although the system
appears to be a three-dimensional network of structural dimers, it
is well possible that there is one dominant interdimer exchange path
which will effectively produce a lower magnetic dimensionality which
can be modeled by an alternating spin chain behavior.

The second approach consists of a mean-field modification of the
Bleaney-Bowers equation
\begin{equation}\label{BBplus}
 \chi(T)=\frac{Ng^2\mu_B^2}{k_B T}[3+ \exp{(J/k_BT)+J^\prime/k_B T}]^{-1},
\end{equation}
where $J^\prime$ is an effective coupling between Cu ions of two
different magnetic dimers.\cite{Nakajima06}

\begin{figure}[h]
\centering
\includegraphics[width=80mm,clip,angle=0]{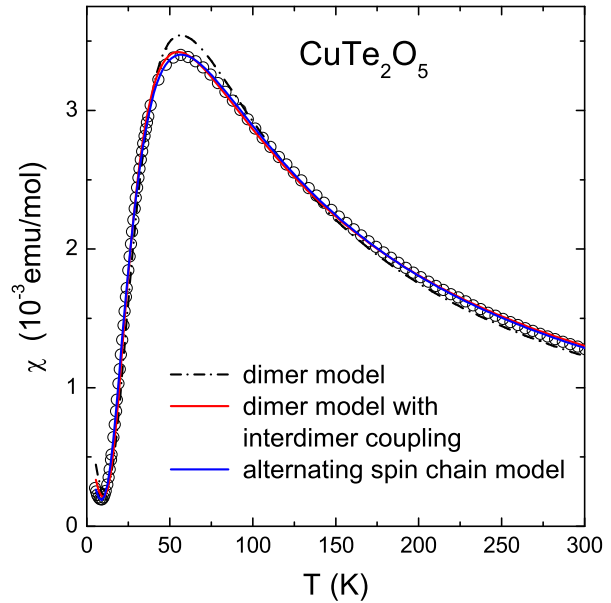}
\vspace{2mm} \caption[]{\label{intensity} (Color online) Temperature
dependences of the spin susceptibility (circles) in CuTe$_2$O$_5$.
The lines are best fits obtained by the models described in the
text.}
\end{figure}

With the above models we tried to describe the spin susceptibility
by obtaining the best fit results which are shown in
Fig.~\ref{intensity}. Note that the very good agreement between the
dc-susceptibility and the ESR intensity allowed us to set
$\chi_{VV}=0$.  Given the existence of structural dimers, we started
to approximate the spin susceptibility by using $\alpha=0$
corresponding to the simple dimer model by Bleaney and Bowers. The
resulting best-fit curve is shown as the dash-dotted line in
Fig.~\ref{intensity}, yielding $J=91.6$~K and $g=1.92$ and
$C=2.44\cdot 10^{-3}$ emu K/mol. Obviously, a pure dimer model fails
to describe the susceptibility in the full temperature range and
yields an effective $g$-factor considerably lower than the one
obtained by ESR measurements.

Varying the ratio $\alpha$ freely, we were able to get an excellent
fit (solid line in Fig.\ref{intensity}) yielding the alternating
exchange coupling constants $J_1=93.3$~K and $J_2=40.7$~K, i.e.~a
ratio $\alpha=J_2/J_1=0.436$, $C=1.45\cdot 10^{-3}$~emu K/mol, and
an effective $g$-factor of $g=1.99$, which is already in reasonable
agreement with the values observed by ESR.

Using Eq.~\ref{BBplus} we can get an equally good fit with
parameters $J=88.9$~K, $J^\prime=91.4$~K, $C=1.84\cdot 10^{-3}$~emu
K/mol and a $g$-factor $g=2.08$ which is in very good agreement with
the experimentally obtained ones. Note that the effective
inter-dimer exchange $J^\prime$ is slightly larger than the
intra-dimer coupling yielding a ratio of $J^\prime/J=1.03$. It was
shown by Sasago et al.\cite{Sasago95} that such the mean-field
approach does not allow to estimate the correct value of $J^\prime$
in the case $J^\prime/J\simeq 1$. Therefore, the above values
obtained by the mean-field equation have to be interpreted with
caution, only allowing to state that the interdimer coupling is
certainly not negligible in this case.

Although we cannot unambiguously determine the magnetic structure by
fitting the susceptibility, we can conclude that both models, the
alternating spin-chain and the modified BB approach, show that in
CuTe$_2$O$_5$ the inter-dimer exchange coupling is of the same order
of magnitude than the intra-dimer coupling. In addition to the
difficulties in describing the ESR data by purely structural
considerations, this finding corroborates that there exists at least
one strong additional magnetic coupling between Cu ions which are
separated by an intermediate Te ligand. To understand the appearance
of such a considerable inter-dimer exchange, it is necessary to
investigate in detail the possible exchange paths between adjacent
Cu ions, which will be subject of the following section.

\subsection{Analysis of the exchange paths}

\begin{figure}[t]
\centering
\includegraphics[width=80mm,clip,angle=0]{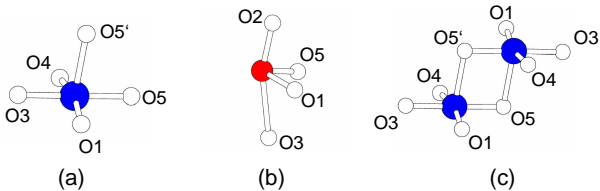}
\vspace{2mm} \caption[]{\label{figure1} (Color online) Building
units of CuTe$_2$O$_5$: (a) CuO$_5$, (b) TeO$_4$ and (c)
Cu$_2$O$_8$. The large, middle and small spheres show Cu, Te and O
atoms, respectively.}
\end{figure}

\begin{figure}[t]
\centering
\includegraphics[width=80mm,clip,angle=0]{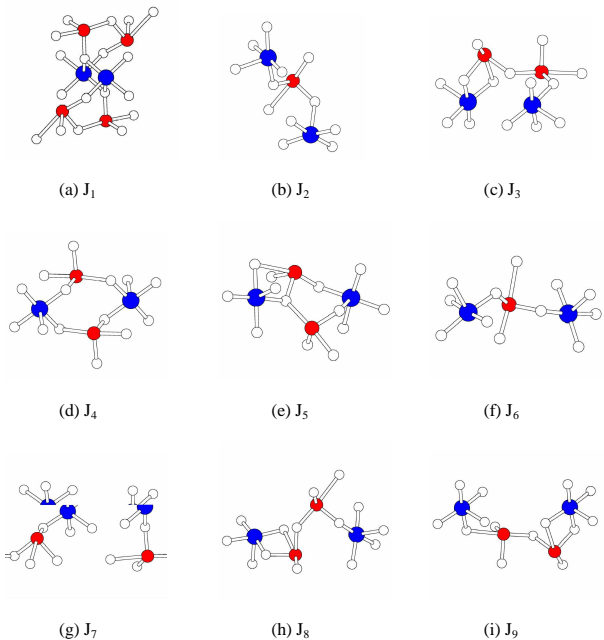}
\vspace{2mm} \caption[]{\label{figure2} (Color online) Spin dimers
of CuTe$_2$O$_5$ associated with the spin exchange interactions
$J_1$-$J_9$.  The large, middle and small spheres show Cu, Te and O
atoms, respectively.}
\end{figure}

Starting out again from the crystal structure of CuTe$_2$O$_5$, we
regard the system as being built up of CuO$_5$ square pyramids
(Fig.~\ref{figure1}a), which edge-share to form structural
Cu$_2$O$_8$ dimers (Fig.~\ref{figure1}c). These dimers corner-share
with TeO$_4$ units (Fig.~\ref{figure1}b) to form the
three-dimensional lattice of CuTe$_2$O$_5$. The spin exchange
interaction within the Cu$_2$O$_8$ dimer is a superexchange (SE)
interaction since its two Cu$^{2+}$ ions are linked by Cu-O-Cu
bridges. Each Cu$^{2+}$ ion of a CuO$_5$ square pyramid (i.e.~a spin
monomer) has one singly occupied $d_{x^2-y^2}$-block orbital
(i.e.~the magnetic orbital) that lies in the basal plane of the
square pyramid. In a Cu$_2$O$_8$ dimer (Fig.~\ref{figure1}c) the
basal planes of the two CuO$_5$ square pyramids are parallel to each
other, so that their magnetic orbitals cannot overlap well.
Therefore, the interpretation of the structural dimer Cu$_2$O$_8$ as
a magnetic dimer may not be justified. However, so far we have not
taken into consideration the role of the TeO$_4$ units linking
between CuO$_5$ square pyramids. Accounting for such TeO$_4$ units
has been shown to change considerably the magnetic properties of
Cu$_4$Te$_5$O$_{12}$Cl$_{4}$ with respect to
Cu$_2$Te$_2$O$_{5}$Br$_{2}$.\cite{takagi06} In general, the spin
exchange interactions of CuTe$_2$O$_5$ can take place through SE
paths, Cu-O-Cu, or through super-superexchange (SSE) paths,
Cu-O...O-Cu, in which the O...O contact is provided by TeO$_4$
units. It has been shown that SSE interactions can be strong in
magnitude and can be even stronger than SE interactions, and hence
must not be neglected.\cite{Whangbo03} Therefore, we evaluated the
relative strengths of nine spin exchange interactions (shown in
Fig.~\ref{figure2}) in CuTe$_2$O$_5$ by performing a spin dimer
analysis based on extended H\"{u}ckel tight binding (EHTB) electronic
structure calculations.\cite{Hoffmann63,Samoa,Whangbo03}

\begin{table}[t] \label{table1}
\caption[]{ Exponents  $\zeta_i$ and valence shell ionization
potentials $H_{ii}$ of Slater-type orbitals $\phi_i$ used for
extended H\"{u}ckel tight-binding calculation. $H_{ii}$ are the diagonal
matrix elements $\langle\phi_i|H_{eff}| \phi_i\rangle$, where
$H_{eff}$ is the effective Hamiltonian. For the calculation of the
off-diagonal matrix elements $H_{ij}=\langle\phi_i|H_{eff}|
\phi_j\rangle$, the weighted formula as described in
Ref.~\onlinecite{Ammeter78} was used.  $C$ and $C^\prime$ denote the
contraction coefficients used in the double-zeta Slater-type
orbital.}

\quad

\begin{tabular}{c|c|c|c|c|c|c}
  atom \, & \,$\phi_i$ \, & \,$H_{ii}$(eV)\, & \,$\zeta_i$\, &\, C \,& \,$\zeta_i^\prime$\, & \,$C^\prime$ \\
  \hline
  Cu & 4s & -11.4 & 2.151 & 1.0 &  &  \\
  Cu & 4p & -6.06 & 1.370 & 1.0 &  &  \\
  Cu & 3d & -14.0 &\, 7.025\, &\, 0.4473\, & \,3.004\, & \,0.6978 \\
  Te & 5s & -20.8 & 4.406 & 0.6568 & 1.652 & 0.4892 \\
  Te & 5p & -13.2 & 3.832 & 0.5934 & 2.187 & 0.5402 \\
  O  & 2s & -32.3 & 2.688 & 0.7076 & 1.675 & 0.3745 \\
  O  & 2p & -14.8 & 3.694 & 0.3322 & 1.866 & 0.7448 \\
  \hline

\end{tabular}
\end{table}

The two magnetic orbitals of a spin dimer interact to give rise to
an energy split $\Delta e$. In the spin dimer analysis based on EHTB
calculations, the strength of an antiferromagnetic interaction
between two spin sites is estimated by considering the
antiferromagnetic spin exchange parameter $J_{AF}=-(\Delta
e)^2/U_{eff}$,\cite{Whangbo03} where $U_{eff}$ is the effective
on-site repulsion that is essentially a constant for a given
compound. Therefore, the trend in the $J_{AF}$ values is determined
by that in the corresponding values $(\Delta e)^2$.

\begin{table}[b] \label{table2}
\caption[]{ Cu$\dots$Cu distances in {\AA},  $(\Delta e)^2$ in (meV)$^2$
for the spin exchange paths $J_1$-$J_9$ in CuTe$_2$O$_5$ shown in
Fig.~\ref{figure2}, and the relative strengths of the spin exchange
interactions compared to the strongest interaction $J_6$.}

\quad

\begin{tabular}{c|c|c|c}
  exchange $J_i$ \, & \,Cu...Cu distance \, & \,$(\Delta e)^2$\,&\,$(\Delta e)^2_i/(\Delta e)^2_6$ \,\\
  \hline
  $J_1$ & 3.187 & 2250 & 0.59 \\
  $J_2$ & 5.282 & 200 & 0.05\\
  $J_3$ & 5.322 & 520 & 0.14 \\
  $J_4$ & 5.585 & 410 & 0.11 \\
  $J_5$ & 5.831 & 40 & 0.01 \\
  $J_6$ & 6.602 & 3840 & 1.00 \\
  $J_7$ & 6.437 & 190 & 0.05 \\
  $J_8$ & 6.489 & 350 & 0.09 \\
  $J_9$ & 6.871 & 990 & 0.26\\
  \hline
\end{tabular}
\end{table}

The magnetic properties of a variety of magnetic solids are well
described by the values obtained from EHTB
calculations,\cite{Whangbo03} when both the $d$-orbitals of the
transition metal and the $s/p$ orbitals of its surrounding ligands
are represented by double-zeta Slater type
orbitals.\cite{Clementi74}
The atomic parameters used for the present EHTB calculations of the
$(\Delta e)^2$ values are summarized in Table~I. The exponents
$\zeta$ and $\zeta^\prime$ of the Slater type atomic orbitals Te
5s/5p listed in Table~I are greater than those of the atomic orbital
calculations\cite{Clementi74} by the factor of 1.45, because such
values are needed in reproducing the band gap of $\alpha$-TeO$_2$ by
EHTB calculations.\cite{Menendez-Proupin04}

The results of our examination of the spin dimers (i.e.the
structural units containing two spin sites) in CuTe$_2$O$_5$ are
listed in Table~II. We identified the strongest interaction $J_6$ to
be of antiferromagnetic SSE type mediated by O-Te-O bridges and the
second strongest $J_1$ to be the antiferromagnetic SE interaction
within the structural dimer Cu$_2$O$_8$, yielding a ratio $(\Delta
e_1)^2/(\Delta e_6)^2=J_1/J_6 = 0.59$. In terms of the description
as an alternating spin chain this ratio is not too far away from the
value $\alpha=0.436$ resulting from the fit with Eq.~\ref{Johnston}.
Therefore, the model of an alternating spin chain appears as the
simplest possible model for CuTe$_2$O$_5$. The existence of further
exchange paths such as $J_3$ and $J_9$ shows, however, that these
spin chains would exhibit non-negligible couplings between each
other. To unambiguously identify the magnetic structure in
CuTe$_2$O$_5$ remains a challenging task for the future, including
the experimental quest for further evidence to corroborate or
discard the possibility of alternating spin chains as suggested by
fitting the spin susceptibility.

The strength of a SSE interaction through the exchange path of the
type Cu-O-L-O-Cu (e.g., L = Te, Se, P) depends sensitively on how
the O-L-O linkage orients the two magnetic orbitals (i.e., the
$d_{x^2-y^2}$ orbital orbitals) of the two Cu$^{2+}$ sites and also
on how the orbitals of the ligand atom L are oriented with respect
to the two magnetic orbitals.\cite{Whangbo03} In
Cu$_2$Te$_2$O$_5$X$_2$ (X=Cl,Br), two adjacent Cu$_4$ tetrahedra
make Cu-O-Te-O-Te-O-Cu and Cu-X$\dots$X-Cu linkages, and it is the
Cu-X$\dots$X-Cu paths that provide strong spin exchange
interactions.\cite{Whangbo03a} In CuSe$_2$O$_5$ the Cu$^{2+}$ ions
form chains of Cu-O-Se-O-Cu linkages leading to a uniform linear
spin chain model.\cite{Kahn80} In CuTe$_2$O$_5$ the spin exchange
$J_6$ has a Cu-O-Te-O-Cu linkage that is quite symmetrical in shape,
so the orbitals of the intervening TeO$_2$ unit provide a strong
overlap with two $d_{x^2-y^2}$ orbitals of the two Cu$^{2+}$ sites.

In general, our finding shows that the intuitive picture of the
equivalence of structural and magnetic dimers can be misleading, as
has been observed for a number of magnetic solids. For example, the
strongest magnetic coupling in CaV$_4$O$_9$  appears between
next-nearest-neighbor vanadium ions,\cite{Pickett99,Koo00} in
contrast to the structurally suggestive picture of only weakly
coupled plaquettes of V$^{4+}$ ions. The singlet-forming Cu$^{2+}$
ions in CaCuGe$_2$O$_6$ are given by the third-nearest-neighbor
copper pairs that occur between chains of edge-sharing distorted
CuO6 octahedra.\cite{Valenti02,Koo05} Both alternating chain and
spin ladder models describe the magnetic susceptibility of
(VO)$_2$P$_2$O$_7$ well,\cite{Johnston87,Barnes94,Johnston01} but an
alternating chain model was proven to be correct for
(VO)$_2$P$_2$O$_7$ by neutron scattering experiments\cite{Garret97}
on oriented single crystals and also by spin dimer analysis based on
EHTB calculations.\cite{Koo02} All these examples demonstrate the
importance of considering the overlap between magnetic orbitals in
arriving at a correct spin lattice model.

\section{Conclusions}

Using electron spin resonance and susceptibility measurements we
determined the spin susceptibility of the Cu-dimer system
CuTe$_2$O$_5$. The obtained ESR data suggested that the structural
dimers do not coincide with the magnetic dimers. The analysis of the
spin susceptibility revealed a coupling of about 90~K within the
magnetic dimer and a considerable inter-dimer coupling of the same
order of magnitude. A detailed investigation of the magnetic
exchange paths revealed the strongest magnetic coupling to be of
super-superexchange type. Therefore, CuTe$_2$O$_5$ belongs to the
interesting class of compounds like CaCuGe$_2$O$_6$ or
(VO)$_2$P$_2$O$_7$, where the effective low energy Hamiltonian
cannot simply be mapped on geometric aspects of the crystallographic
structure. Eventually, such a discrepancy might be found in many
low-dimensional compounds and it appears a challenging experimental
and theoretical task to identify and classify common parameters of
such systems.

\begin{acknowledgments}
It is a pleasure to thank D.~Zakharov for fruitful discussions. We
acknowledge support by the BMBF via contract number VDI/EKM 13N6917
and partly by the DFG via SFB 484 (Augsburg), the Swiss NSF through
the NCCR MaNEP, and by the RFBR (Grant No. 06-02-17401) and BRHE REC
007. The research was further supported by the Office of Basic
Energy Sciences, Division of Materials Sciences, U. S. Department of
Energy, under Grant DE-FG02-86ER45259. K.-S. L. thanks The Catholic
University of Korea for the 2006-Research Fund.
\end{acknowledgments}

\end{document}